\documentclass[twocolumn]{aa}
\usepackage{graphicx}
\usepackage{txfonts}
\begin{document}
   \title{In flight calibration of the ISGRI camera}

   \author{
          R. Terrier
          \inst{1}
	  \and
          F. Lebrun
	  \inst{1}
	  \and
	  A. Bazzano
	  \inst{2}
	  \and
	  G. Belanger
	  \inst{1}
	  \and
	  A.J. Bird
	  \inst{5}
	  \and
          C. Blondel
	  \inst{1}
	  \and  
	  P. David
	  \inst{1}
	  \and
	  P. Goldoni
	  \inst{1}
	  \and
	  A. Goldwurm
	  \inst{1}
	  \and
	  A. Gros
	  \inst{1}
	  \and
	  P. Laurent
	  \inst{1}
	  \and
	  G. Malaguti
	  \inst{4}
	  \and
          A. Sauvageon
	  \inst{1}
	  \and
          A. Segreto
	  \inst{3}
	  \and
	  P. Ubertini
	  \inst{2}
}

   \offprints{R. Terrier, rterrier@cea.fr}

   \institute{
     Service d'Astrophysique, CEA/Saclay, France
\and
     Insituto di Astrofisica Spaziale  e Fisica cosmica, CNR, Rome, Italy
\and
     Insituto di Astrofisica Spaziale  e Fisica cosmica, CNR, Sezione di Palermo, Italy
\and
     Insituto di Astrofisica Spaziale  e Fisica cosmica, CNR, Sezione di Bologna, Italy
\and
     School of Physics and Astronomy, University of Southampton, Southampton, UK 
             }


   \abstract{ ISGRI, the IBIS low energy camera (15 keV - 1 MeV) on board INTEGRAL,  is the first large CdTe gamma-ray imager in orbit.   
 We present here an overview of the ISGRI in-flight calibrations performed during the first months after launch. We discuss the stability of the camera as well as the CdTe pixels response under cosmic radiation. The energy calibrations were done using lead and tungsten fluorescence lines and the $\mathrm{^{22}Na}$ calibration unit. Thermal effects and charge correction algorithm are discussed, and the resulting energy resolution is presented. The ISGRI background spatial and spectral non-uniformity is also described, and some image correction results are presented.
                
   \keywords{INTEGRAL, gamma-rays, calibration
               }
   }

   \maketitle
%

\section{Introduction}

The ISGRI low energy camera (15 keV-1 MeV) of the IBIS imager (\cite{Ubertini03}) on-board INTEGRAL (\cite{Winkler03}) is an array of 128 by 128  Cadmium Telluride (CdTe) pixels $4\times 4\times 2$ $\mathrm{mm^3}$ in size. They are grouped in 8 Modular Detector Units (MDU) which are assembled on an aluminium structure. For more details on the design and expected performances, see \cite{Lebrun03}.
CdTe has a large photoelectric cross section, due to its high Z. However, ballistic losses cause an incomplete collection of the charges created during the gamma-ray interaction and thus produce a large continuum up to the photopeak energy. Nevertheless since the charge loss is proportional to the transit time which governs the pulse rise time, the deposited energy can be corrected using a measurement of rise time. 
For this reason, the ISGRI ASIC has been designed to read both the pulse height and rise time of 4 pixels, and a charge correction algorithm has been developed to provide  good energy resolution over the whole energy range (\cite{Lebrun96}). The values obtained on ground vary from 8\% at 60 keV to 4\% at 511 keV.

The in-flight calibrations have been performed using scientific data from Cygnus X-1, Crab, Galactic Center and empty field observations, as well as calibration unit data (in coincidence with the $\mathrm{^{22}Na}$ source on board also called S2 data, see \cite{Ubertini03} and \cite{Bird}). The aim is to:
\begin{itemize}
\item provide a good algorithm to detect and suppress bad and noisy pixels
\item update the pulse height and rise time gains and offsets determined on ground using in-flight background and calibration source lines
\item check the validity and update the charge loss correction algorithm
\item determine the background structure and provide a correction for deconvolution
\end{itemize}
Calibration of the ISGRI point spread function (PSF) (\cite{Gros03}) as well as the instrument response matrix modeling and  in-flight validation (\cite{Laurent03}) are presented elsewhere.

\section{Pixel behaviour}

When ISGRI was first switched on, the pixels were randomly bursting with events having a null rise time. The origin of these events is still unclear. They represent around 15\% of the total ISGRI count rate, and since they are useless, a patch was applied to the on-board software to reject low rise time events. 
Moreover,  noisy pixels are cut on-board by the NPHS (Noisy Pixels Handling System), which detects high number of triggers in a single pixel and in a module. Because of the zero rise time events, it was necessary to lower the NPHS sensitivity in order to limit the number of pixels switched off. For more details on NPHS and the zero rise time events, see also \cite{Lebrun03}.

To limit the number of noisy pixels,  a ground based  algorithm has been developed to update the thresholds of each pixel depending on its individual spectrum after every revolution.
Noisy pixels have a strong low energy peak but have a correct high energy spectrum. A new threshold is computed in order to remove the whole peak. Bad pixels usually have a low efficiency and a very hard spectrum with a large bump between 50  and few hundred keV. They are selected using an hardness ratio and switched off. All the pixels, identified as noisy in a previous revolution, have their threshold lowered by one step ($\sim$ 1 keV).

\begin{figure}
  \centering
  \includegraphics[width=8.5cm]{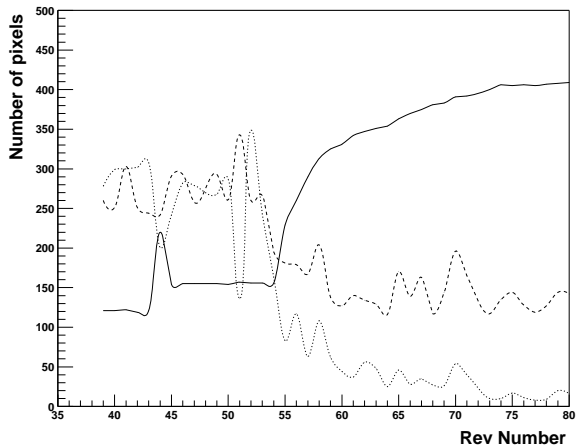}
      \caption{Evolution of the number of disabled ISGRI pixels  (black solid line), and number of pixels which efficiency is higher (dashed line) and less (dotted line) than 70\%, from revolution 35 (starting on 01/21/03) to 80 (ending on 06/12/03). The introduction  of the new pixel thresholding algorithm during revolution 54 removed progressively most of the bad pixels.}
         \label{FigPixelLoss}
   \end{figure}

The introduction of the aforementioned algorithm has strongly increased the number of pixels switched off and is now steady at around 420, i.e. less than 3\% of the camera. In the same time the number of low efficiency pixels has decreased, as can be seen on figure \ref{FigPixelLoss}. The number of noisy pixels found in each revolution is $\sim$250. The total fraction of pixels switched off or noisy is therefore $\sim$4.5\%, which is better than the 5\% specification.

Some other unexpected features of the CdTe pixels response in flight can be seen on the background biparametric diagram on figure \ref{FigBipar}. For instance, a large horizontal structure at high rise time can be seen spanning over the whole energy range. Its origin is still unclear, though likely due to cosmic rays since it was barely visible on ground tests. It does not strongly limit the sensitivity since there are not many events at such high rise times. An on board upper limit has been applied to remove those background events and lower the telemetry load.


\section{Energy calibration}

The gains and offsets of rise time and pulse height have been calibrated on the ground at various biases and temperatures using several sources. A look-up table (LUT1) valid at $\mathrm{0^{\circ}C}$ and 100V was produced, and a dependency on temperature and bias was found. This is detailed in the next section and in \cite{Lebrun03}. 

In order to check the validity of the LUT1, the background lines were used to calibrate the in-flight gains and offsets. It is important to note that the calibration of rise time is especially difficult because there are no special features in the rise time distribution. Therefore, it is necessary to build a biparametric diagram for the 511 keV line for each pixel, but this requires a much larger amount of data than is currently available.

\begin{figure}
  \centering
  \includegraphics[width=8.5cm]{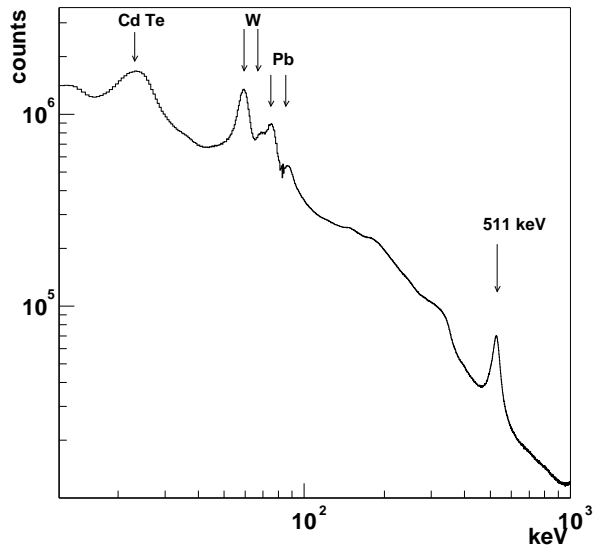}
      \caption{ISGRI in-flight calibration data spectrum, after thermal and charge loss corrections. Note the large line at around 25 keV and the four lead and tungsten fluorescence lines.}
      \label{FigSpec}
\end{figure}

The calibration data spectrum, shown on figure \ref{FigSpec} is characterised by several lines: the fluorescence lines of tungsten (59.3 keV and 67.2 keV) and  lead (75 keV and 84.9 keV), and also the 511 keV line a portion of which is due to $\mathrm{^{22}Na}$ calibration source (see \cite{Bird}). These is also a large peak at low energies (around 24 keV) whose origin is not completly understood. 
The fluorescence lines of Cd (23.1 and 26.9 keV) and Te (27.5 and 31 keV) certainly contribute.  Besides, the irradiation of CdTe creates a large number of radioactive isotopes among which $\mathrm{^{111m}Cd}$ could contribute (\cite{Murakami03}).  It has a life time of 45 minutes, and decay with the emission of a 150 keV line (which is marginaly detected because of the reduced peak efficiency), and also lines between 23 and 26 keV.

\subsection{Thermal effects}

The temperature dependence of gains and offsets has been evaluated on ground for temperatures ranging from $\mathrm{-10^{\circ}C}$ to $\mathrm{20^{\circ}C}$. It was found that the offsets are roughly constant, but that both pulse height and rise time gains strongly depend on temperature. They were fitted with the following expressions:
\[
{\frac{g_{\mathrm{PH}}}{g_{\mathrm{PH}}(273\mathrm{K},100\mathrm{V})} =  \left(\frac{T}{273\mathrm{K}}\right)^{-1.11} \left(\frac{U}{100 \mathrm{V}}\right)^{-0.08}}
\]
\[
{\frac{g_{\mathrm{RT}}}{g_{\mathrm{RT}}(273\mathrm{K},100\mathrm{V})} =  \left(\frac{T}{273\mathrm{K}}\right)^{0.52} \left(\frac{U}{100 \mathrm{V}}\right)^{0.58}}
\]
$g_{\mathrm{PH}}$ and $g_{\mathrm{RT}}$ being expressed respectively in keV per channel and $\mathrm{\mu s}$ per channel.

\begin{figure}
  \centering
  \includegraphics[width=8.5cm]{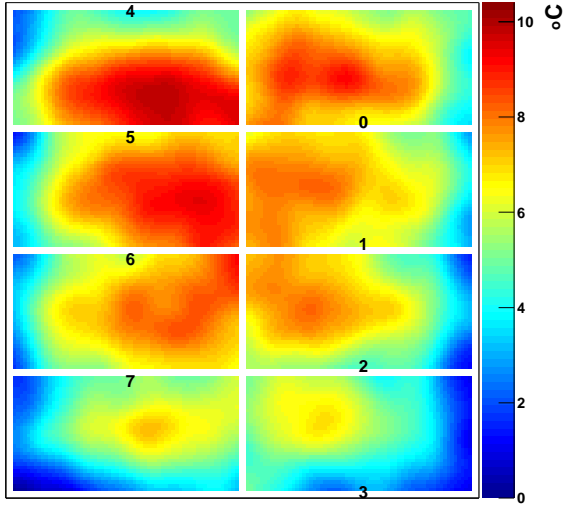}
      \caption{Effective temperature map on the camera: relative temperature variation between the thermal sensor number 3 and the temperature deduced from the pixel gains. A module per module structure is clearly visible, as well as a thermal gradient along the z-axis. Numbers indicates MDU thermal sensors positions in the camera.}
      \label{FigTemp}
   \end{figure}

 The temperature variations can be as large as 10K in a single revolution, if the pointing is changed or some PICSIT modules are switched off in the radiation belts. This can induce gain variations of the order of 5\% which can be strong limiting factors for the energy resolution especially at high energies where the relative resolution is significantly lower. Thermal sensors are placed on the side of every ISGRI module as shown on figure \ref{FigTemp} and allow us to correct for temperature effects on a module per module basis. 

The pulse height gains and offsets were then calibrated using the 59.3 keV fluorescence line, and the 511 keV line in S2 data. A systematic bias was observed between the LUT1 and the coefficients obtained. This bias can be interpreted in terms of temperature variations in the camera, for the on-ground calibration was not done in vacuum; the thermal equilibrium is therefore very different. Using the previous formula, an effective temperature map was deduced, and is shown in figure \ref{FigTemp}. Once the temperature is taken into account, the pulse height calibration errors are very small. Note also that this correction is the only one applied to the rise time  gains and offsets.

\subsection{Charge loss corrections, LUT2}

Charge losses have been measured on ground using various radioactive sources: $\mathrm{^{22}Na}$,  $\mathrm{^{139}Ce}$,  $\mathrm{^{54}Mn}$,  $\mathrm{^{57}Co}$,  $\mathrm{^{133}Ba}$, etc. The pulse height -- rise time relations obtained for various energies  were interpolated to produce a general charge correction look-up table (LUT2). This is detailed in \cite{Lebrun03}. 

\begin{figure*}
  \centering
  \includegraphics[width=15cm]{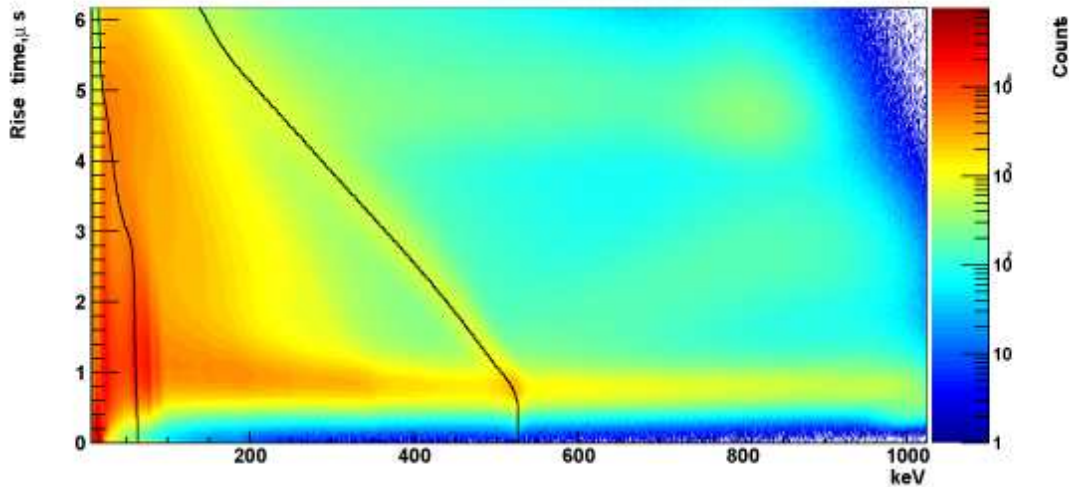}
  \caption{In flight biparametric diagram obtained with calibration unit data. Superimposed is the charge correction dependency from the LUT2 for the 59.3 keV and 511 keV lines. The background displays an unexpected high rise time structure covering the whole amplitude range. This strongly limits the sensitivity at high rise times, therefore  an on-board upper limit selection has been implemented. A second unexpected structure can also be seen with pulse height increasing with rise time. The origin of these structures is unclear.}
  \label{FigBipar}%
\end{figure*}

The in-flight bias voltage and temperature conditions being different from those used for the LUT2 production, a simple correction was applied to it in order to recover good performances for the 59.3 and 511 keV lines: a multiplicative factor and an offset have been applied to the LUT2 so that both lines match those observed in the biparametric diagram of the flight calibration data (see figure \ref{FigBipar}).

\subsection{Energy resolution}

\begin{figure}
  \centering
  \includegraphics[width=8.5cm]{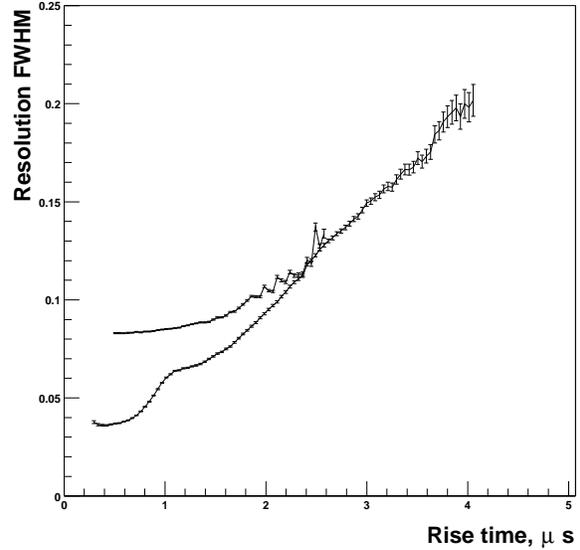}
      \caption{Variation of energy resolution (FWHM) with rise time at 59.3 keV and 511 keV. The strong degradation with rise times comes from the remaining miscalibations of rise time gains and offset. Note that the 59.3 keV resolution is not given above 2.6 $\mathrm{ \mu s}$  because the tungsten $K\alpha$ line is mixed up with $K\beta$ and lead fluorescence lines.}
        \label{FigRes}
\end{figure}

After correcting for temperature effects, we evaluated the energy resolution for the 59.3 keV and 511 keV lines using S2 data, and studied its variation with rise time.  The resolution obtained for low rise time is coherent with expectations, but it increases faster with rise time as is shown on figure \ref{FigRes}. At 511 keV, the integrated width (up to 4 $\mu s$) is of the order of 25 keV (FWHM) resulting in a 4.9\% resolution, which is larger than what was expected from ground measurements. This is primarly caused by the rapid degradation of resolution with rise time: from 3.5\%  to more than 20\% at $\mathrm{4 \mu s}$. Poor calibration of rise time gain and offset is mainly responsible for the degradation, and an effort is needed in order to improve these performances. 
This effect is less important at lower energies where the charge losses are less important.  At 59.3 keV, the integrated FWHM is 5 keV resulting in a 8.4\% resolution. In this case, the variation with rise time is small, since charges losses are low, see figure \ref{FigRes}.

\section{Instrumental background}

The background shape and spectrum is a key issue in optimizing the sensitivity of a coded mask imaging system like IBIS. The various contributions to the background include:  galactic and extragalactic diffuse gamma-ray emission that dominates in the low energy range, nuclear lines produced in the external payload and spacecraft mass distribution, as well as scattering of cosmic rays. In the standard analysis (\cite{Goldwurm03}) a background model  is subtracted from the shadowgram before the deconvolution is performed. This background model is produced using empty fields observations or by summing shadowgrams obtained in dithering mode. For more details on the background non-uniformity in general, see \cite{Natalucci03}.

We have studied the background spatial and spectral behaviour for two different VETO configurations: using only lateral anticoincidence (veto LAT) up to revolution 33, and with lateral and bottom coincidence (veto ALL) for subsequent revolutions. To build a background model, we have used two empty field staring observations (rev 14 and 24) and a part of revolution 38 as well as some high latitude observations where the source contribution is weak (veto ALL). Images were produced with these data in various energy bands correcting for bad and noisy pixels as well as pixel efficiency in each science window. These images are then used as background model, leaving the normalization parameter free, because the background structures are supposed to be constant, but the global flux is expected to be variable. 

\begin{figure*}
  \centering
  \includegraphics[width=18cm]{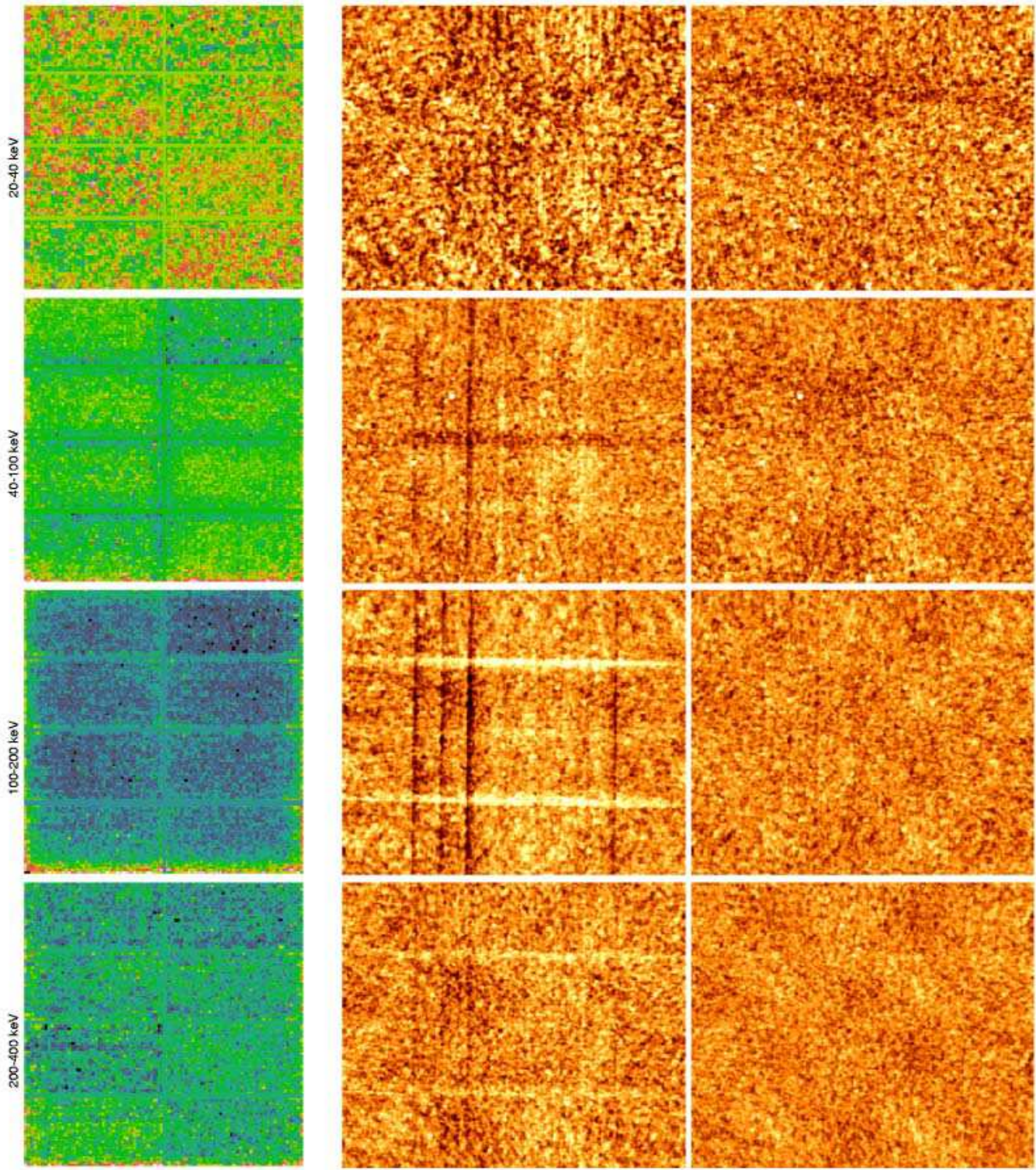}
      \caption{Effect of noise substraction on mosaic images in various energy bands (20-40keV, 40-100 keV, 100-200 keV, 200-400 keV). First column show the background model shadowgram, second column the mosaic of significance obtained without background substraction and third column the same mosaic after substraction. Significance values range from $-5\sigma$ (black) to $5\sigma$ (white). Strong background features in particular in intermediate energy bands are clearly removed. Some residual structure is still visible and is currently under investigation.
}
         \label{FigBack}
\end{figure*}

Figure \ref{FigBack} (left column) shows the background model obtained in several standard imaging energy bands in the vetoALL configuration. The significance mosaic maps obtained with the standard analysis for three science windows with (figure \ref{FigBack}, middle) and without (figure \ref{FigBack}, right) correction are also shown. In the low energy regime ($<$40 keV), a rather flat structure is observed in the camera. The shadow of the aluminium frame is visible on the first pixel row in the upper and lower modules, suggesting an important contribution of diffuse gamma-rays to the background. From 40 to 100 keV, the background is dominated by the contribution of tungsten and lead fluorescence lines coming from the hopper and the mask. Since the incidence angle on the camera can be large, the aluminium frame's shadow is much broader, thus creating convex module structures. At higher energies (see figure \ref{FigBack}, 100-200 and 200-400 keV bands), the background is more intense on the sides of the camera, especially on the lower side closest to SPI, due to scattering in passive materials and possible leaks in the shielding. The pixels around the aluminium frame also show a strong flux enhancement for the same reasons, in this case creating a concave module structure. These structures produce very strong lines in the deconvolved images. In the vetoLAT configuration, the high energy regime is also characterized by a large bump in the whole detector producing a saddle shape in the resulting images.

Subtracting the background model from the shadowgrams yields flat sky maps, as shown on the right column of figure \ref{FigBack} which was obtained  using 3 Galactic center observations science windows. The bright lines, due to background enhancement on the detector sides, seen above 100 keV are very well suppressed, but some errors remain although at a much smaller level.
However, the background variations with time are large and the model used does not correct properly all observations, depending on latitude, cosmic ray fluxes, and status of the camera. The background structure variations are under study in order to quantify these as a function of time, latitude and cosmic ray flux.

A final point is that background substracted images have to be corrected for efficiency non-uniformity due to for instance gamma-ray scattering (see \cite{Natalucci03}). This, however, is a second order effect, and it is not taken into account at the moment.   

\section{Conclusion}
ISGRI is the first large CdTe gamma-ray imager used in orbit, and despite a few unexpected features like zero rise time events, it performs very well with only 4.5\%  noisy or disabled pixels. The energy calibration has been performed using the on board calibration source and background lines. Thermal effects are at the origin of the largest difference between ground and in-flight data. Correcting for these effects yields good spectral performances close to the expectations with 8.4\% at 59.3 keV and 4.9\% at 511 keV. The resolution in the high energy band is broader than before launch because of residual rise time gains uncertainties. Handling of these errors requires a larger amount of calibration data than what is available today. 

Background non-uniformity in the camera has also been studied, and a simple energy dependent model has been produced using empty field observations. It provides rather good flat fielded images, even above 100 keV where the background non uniformity is largest. Unfortunately, background variations with time are large, and further studies are required in order to attain the full sensitivity.

\begin{acknowledgements}
    R.T. acknowledges financial support from CNES.
    This work is based on observations with INTEGRAL, an ESA project with instruments
    and science data centre funded by ESA member states (especially the PI
    countries: Denmark, France, Germany, Italy, Switzerland, Spain), Czech
    Republic and Poland, and with the participation of Russia and the USA.

\end{acknowledgements}

\end{document}